\documentclass[
  aps,            
  pra,            
  reprint,        
  superscriptaddress, 
  floatfix,       
  showkeys,       
  nofootinbib     
]{revtex4-2}

\usepackage{amsmath,amssymb,bm}   
\usepackage{graphicx}             
\usepackage[colorlinks,citecolor=blue]{hyperref}       
\usepackage{color}                
\usepackage{siunitx}              
\usepackage{braket}               
\usepackage{physics}              
\usepackage{dcolumn}              
\usepackage{comment}              
\usepackage{appendix}
\usepackage{placeins}

\newcommand{\down}{\downarrow}
\newcommand{\up}{\uparrow}

\newcommand*{\affaddr}[1]{#1} 
\newcommand*{\affmark}[1][*]{\textsuperscript{#1}}

\begin{document}
\title{Dynamics of antiferromagnetic Dimers in Rydberg Atom Chains}
\author{
Feng-Yuan Kuang\affmark[1], Lin Li\affmark[1], and Weibin Li\affmark[2]\\
\affaddr{\affmark[1]National Gravitation Laboratory, MOE Key Laboratory of Fundamental Physical Quantities Measurement, Hubei Key Laboratory of Gravitation and Quantum Physics, PGMF, Institute for Quantum Science and Engineering, School of Physics, Huazhong University of Science and Technology, Wuhan 430074, China}\\
\affaddr{\affmark[2]School of Physics and Astronomy, and Centre for the Mathematics and Theoretical Physics of Quantum Non-equilibrium Systems, The University of Nottingham, Nottingham NG7 2RD, United Kingdom}
}

\date{\today}

\begin{abstract}
We investigate the dynamics of antiferromagnetic dimers within a Rydberg atom chain in the regime where laser detuning compensates for nearest-neighbor (NN) interactions. Using an effective PXQ model, we demonstrate that the associated Hilbert space decomposes into disconnected, dimer-conserving subspaces. The classification of these subspaces is provided, and  the computational basis states spanning them are identified. Through a combination of analytical mapping and numerical simulations, we compare the dynamics of the PXQ model with those of the full Rydberg atom chain. The deviations are attributed to two factors, laser-induced leakage from the constrained Hilbert subspace and the influence of long-range interactions beyond the NN limit. Our results indicate that subspace leakage can be mitigated by increasing the NN interaction strength. While this simultaneously amplifies the effects of long-range interactions, the conservation of the dimer number remains.  Our study opens up possibilities for exploring the dynamics of antiferromagnetic dimers using the Rydberg atom quantum simulator.
\end{abstract}

\maketitle

\section{Introduction}
Strongly interacting Rydberg atom provide a versatile setting for probing constrained quantum dynamics \cite{Igor2011Many-Body, bernien2017probing, Choi2019Emergent, Khemani2019Signatures, Lin2020Slow, Karle2021Area-Law, Surace2021Exact,  Verresen2021Prediction, Ivanov2025Volume-Entangled, Kerschbaumer2025Quantum, Luke2025Integrable, Soto-Garcia2025Numerical}. The Rydberg blockade enforces local constraints on the Rydberg atom excitation that give rise to the PXP Hamiltonian~\cite{Fendley2004Competing,Igor2012Interacting,Omiya2023Quantum},
which, despite being nonintegrable, hosts a small set of non-thermal eigenstates embedded within an otherwise chaotic spectrum \cite{Turner2018Quantumscarred, Turner2018Quantumscarred, Turner2018Weak, Choi2019Emergent}. This gives the so-called quantum many-body scars. The associated dynamical signatures render the PXP model a canonical example of weak eigenstate thermalization hypothesis (ETH)~\cite{Deutsch1991Quantumstatistical, Srednicki1994Chaos, D'Alessio2016quantumchaos, Deutsch2018Eigenstate} violation in constrained many-body systems~\cite{Ho2019Periodic, Iadecola2019Quantum, Lin2019Exact, Michailidis2020Slow, Surace2020Lattice, Turner2021Correspondence, Rozon2022Constructing, Tomasz2022Unsupervised, Windt2022Squeezing,  Giudici2024Unraveling}. Beyond the Rydberg atom setting, ergodicity breaking dynamics~\cite{Nandkishore2015ManyBody, Turner2018Quantumscarred, Abanin2019Colloquium, Choi2019Emergent, Alhambra2020Revivals, serbyn2021quantum, Mondragon2021Fate, Moudgalya2022quantum, Chandran2023Quantum} can be probed in other systems, such as the superconducting qubit processors~\cite{Wang2025Exploring} and
ultracold atoms~\cite{Su2023Observation,adler2024observation,honda2025observation}. A common feature is that the kinetic constraints~\cite{Shiraishi2017Systematic,Iadecola2020Quantum,Yang2025constructing} and emergent conservation laws~\cite{Khemani2020Localization,Sala2020Ergodicity,Moudgalya2022Hilbert} severely reduce the dynamical connectivity~\cite{Magoni2024Coherent,Yang2025Probing}.

In deriving the PXP model in the Rydberg atom chain, the laser is resonant (or near resonant) with the transition between the ground state and Rydberg state ~\cite{Igor2011Many-Body, Choi2019Emergent, Khemani2019Signatures, Lin2020Slow, Karle2021Area-Law, Surace2021Exact, Verresen2021Prediction, Ljubotina2023Superdiffusive, Su2023Observation, Ivanov2025Volume-Entangled}. 
When the laser detuning is not negligible, the competition among laser detuning,  Rabi frequency, and strong interactions leads to interesting phenomena. For example, a particularly interesting situation is the so-called anti-blockade regime, where the detuning compensates for the Rydberg atom interaction for atoms separated by a given distance~\cite{Ates2007Antiblockade, Pohl2009Breaking, Qian2009Breakdown, Amthor2010Evidence, Su2017Fast, Zhao2017RobustGroundStateAntiblockade, Kara2018RydbergInteractionThermalVapor, Su2021Dipole-dipole,  Zhao2023FloquetTailored}. 
The anti-blockade mechanism has been used to increase two-photon Rydberg excitation probability~\cite{Ates2007Antiblockade, Pohl2009Breaking, Qian2009Breakdown, Amthor2010Evidence}, build
fast quantum gates~\cite{Su2017Fast, Su2021Dipole-dipole}, 
and enhance interaction effects~\cite{liNonadiabaticMotionalEffects2013a, marcuzziFacilitationDynamicsLocalization2017, Zhao2017RobustGroundStateAntiblockade, Kara2018RydbergInteractionThermalVapor, liuLocalizationCriticalityAntiblockaded2022, Zhao2023FloquetTailored}. 
Recent achievements in experimental control have allowed access to constrained models in more general situations~\cite{Desaules2023Prominent, bluvstein2024logical,  chiu2025continuous, gao2025signal, Morettini2025Transport, Ma2025Liouvillian, pizzi2025genuine, wang2025tricritical}. This allows for the observation of distinctive connectivity and unconventional Krylov thermalization with clustered interactions beyond the PXP model~\cite{zhao2025observation}. 
Another intriguing regime arises when the laser detuning equals the nearest-neighbor interaction strength \cite{Igor2017Facilitation,Ostmann2019localization,Van2021disorder}. In this case, the many-body dynamics is governed by an effective PXQ model, in which an atom can change its electronic state only if its two neighboring atoms are in the ground and Rydberg states, respectively.  
\begin{figure}[htbp]
    \centering
    \includegraphics[width=1.0\linewidth]{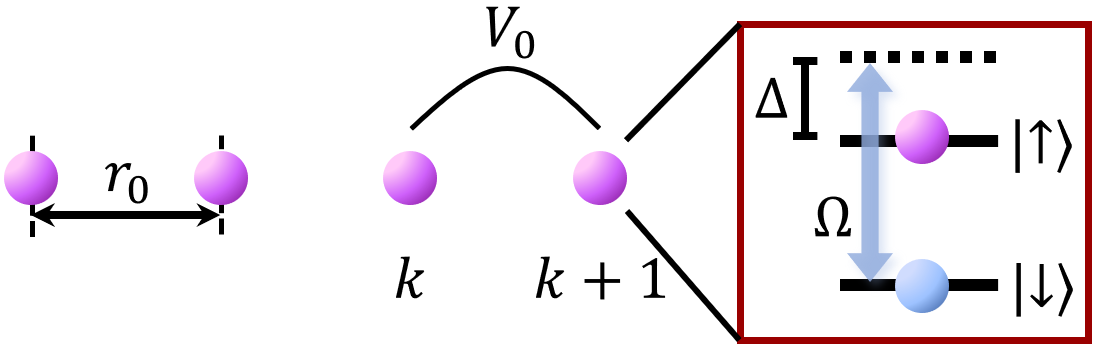}
    \caption{One-dimensional Rydberg atom chain. The neighboring atoms are separated by a distance $r_0$. Atoms are excited from the ground state $\ket{\downarrow}$ to the Rydberg state $\ket{\up}$ with Rabi frequency $\Omega$ and detuning $\Delta$. In the Rydberg state, atoms interact via van der Waals interactions. $V_0$ is the nearest-neighbor interaction term.  
    }
    \label{fig:setting}
\end{figure}

In this study, we investigate properties and dynamics of the antiferromagnetic dimers in a strongly interacting Rydberg atom chain. When the NN interaction compensates for the detuning, the dynamics is approximately described by the PXQ model. By analyzing the connectivity structure of the PXQ Hamiltonian, we systematically characterize the fragmented subspaces that arise under this constraint. We show that the number of antiferromagnetic dimers is a conserved quantity in this model. Beyond the single-excitation subspace studied previously \cite{Igor2017Facilitation,Ostmann2019localization}, we show that the Néel-state subspace exhibits qualitatively distinct connectivity patterns and dynamical behavior. Building on these examples, we establish the general structure and symmetries of the Hilbert subspaces, confirming that the PXQ model can be mapped to the Heisenberg XX model. We show that caution is required when studying dimer dynamics in Rydberg atom chains. Although stronger interactions help suppress fast-rotating terms in the derivation of the effective PXQ model,   the long-range tails of the interaction become non-negligible. Despite this deviation, we show that nontrivial dynamics, such as the conservation of the antiferromagnetic dimer, can be probed in the Rydberg atom chain.

The remainder of this paper is organized as follows. Section \ref{Hamiltonian and the structure} introduces the PXQ Hamiltonian, obtained from the Hamiltonian of the Rydberg atom under single-photon resonance and strong interaction conditions.  The constraints of the PXQ model are then outlined. Section \ref{Properties of the PXQ model} analyzes the properties of the PXQ model, focusing on its connectivity structure, distinctive coupling symmetries, and  classification of the subspaces. Section \ref{Dynamics in fragmental subspaces} explores the dynamics of the representative subspaces. We compare the dynamics of the PXQ model and the Rydberg atom Hamiltonian. The tail of the long-range interaction leads to a deviation from the PXQ model. However, the features of the dimer can still be found in the dynamical evolution of the Rydberg atoms. We conclude in Sec.~\ref{sec:conclusion}. 

\section{Effective dimer Hamiltonians}
\label{Hamiltonian and the structure}

We consider a one-dimensional chain of $L$ Rydberg atoms with an open boundary, as illustrated in Fig.~\ref{fig:setting}. The spacing between two neighboring atoms is $r_0$. The atom is excited from the ground state $\ket{\down}$ to a Rydberg state $\ket{\up}$ by a detuned laser with Rabi frequency $\Omega$ and detuning $\Delta$. In the Rydberg state, atoms interact strongly through the van der Waals (vdW) interaction $V_{jk} = C_\alpha / |\mathbf{r}_j-\mathbf{r}_k|^6$~\cite{saffmanQuantumInformationRydberg2010, adamsRydbergAtomQuantum2019, shaoRydbergSuperatomsArtificial2024a}, with $C_6$ and $\mathbf{r}_j$ to be the dispersion coefficient and location of the $j$-th atom, respectively. The Hamiltonian of the atom chain is given by,
\begin{align*}
\hat{H}_{0} &= \sum_k \left[\frac{\Omega}{2}\hat{\sigma}_k^x - \Delta \hat{Q}_k  + \sum_{j>k} V_{jk}\hat{Q}_j \hat{Q}_k \right], 
\end{align*}
where $\hat{\sigma}_k^x = \ket{\up_k}\bra{\down_k} + \ket{\down_k}\bra{\up_k}$ describes transitions between states $\ket{\down_k}$ and $\ket{\up_k}$ of the $k$-th atom. $\hat{Q}_k = \ket{\up_k}\bra{\up_k}$ and $\hat{P}_k = \ket{\down_k}\bra{\down_k}$  are the projection operators for the Rydberg and ground state, respectively. 

The PXQ model is obtained in the strong interaction regime, characterized by the nearest-neighbor (NN) interaction $V_0=C_6/r_0^6\gg \Omega$. When the detuning equals the NN interaction $\Delta = V_0$, dynamics of the system is described by an effective Hamiltonian (see Appendix~\ref{app:H_e}),
\begin{align}
\hat{H}_{\text{e}}\approx \hat{H}_{\text{PXQ}} +\sum_{|j-k|>1} V_{jk}\hat{Q}_j \hat{Q}_{k} ,\nonumber
\end{align}
where the PXQ Hamiltonian reads,
\begin{align}
\hat{H}_{\text{PXQ}} = \frac{\Omega}{2}\sum_{j=1}^L \left( \hat{P}_{j-1}\hat{\sigma}_{j}^x\hat{Q}_{j+1} + \hat{Q}_{j-1}\hat{\sigma}_{j}^x\hat{P}_{j+1}  \right), \label{eq:Hpxq}
\end{align}
and  $\hat{N}_{\text{cl}} = \sum_k \hat{Q}_k \hat{P}_{k+1}$. 
Operator $\hat{N}_{\text{cl}}$ counts the number of dimers with the antiferromagnetic configuration $\ket{\up\down}$ along the chain. 
It can be shown that $[\hat{N}_{\text{cl}}, \hat{H}_{\text{e}}]=0$ and $[\hat{N}_{\text{cl}},\hat{H}_{\text{PXQ}}]=0$.
In other words, $\hat{N}_{\text{cl}}$ is a conserved quantity, which allows us to study the properties of the antiferromagnetic dimers with both Hamiltonians. Long-range interactions, such as next-nearest-neighbor (NNN) interactions and other terms with separation $|\mathbf{r}_j-\mathbf{r}_k|\ge 2r_0$, are much smaller than the NN interaction, as the vdW interaction decays rapidly with distance. If we neglect the last term in the Hamiltonian $\hat{H}_{\text{e}}$, this leaves the dynamics primarily governed by $\hat{H}_{\text{PXQ}}$. In the following, we discuss the properties of the dimer in Hamiltonian  $\hat{H}_{\text{PXQ}}$. The impact of long-range interactions on the PXQ model is discussed in Section~\ref{Dynamics in fragmental subspaces}.
\begin{figure}
    \centering
    \includegraphics[width=0.7\linewidth]{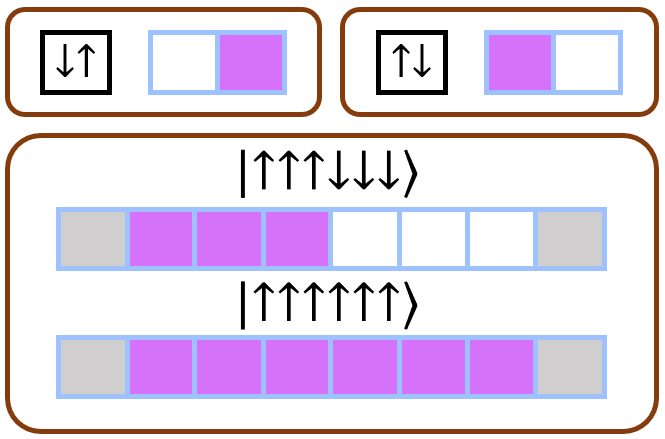}
    \caption{Dimer configurations and boundaries of the chain.  We illustrate dimers with $N_{\text{cl}}=1$ and  $L=6$. Beyond the boundaries, the atoms are in the ground state (grey sites). This ensures that the number of $\ket{\down\up}$ and $\ket{\up\down}$ dimers is equal. Any two neighboring dimers must be separated by a continuous block of excited atoms.}
    \label{fig:cluster}
\end{figure}

\section{Properties of the PXQ model}
\label{Properties of the PXQ model}
\subsection{ \texorpdfstring{Subspace Decomposition with $\hat{N}_{\text{cl}}$}{block decomposition with Ncl}} 

In the PXQ Hamiltonian, the three-site operators couple states that conserve the number of the dimers. For example,  these operators allow the coupling processes $\ket{\uparrow\uparrow\downarrow}\leftrightarrow\ket{\uparrow\downarrow\downarrow}$ and $\ket{\downarrow\uparrow\uparrow}\leftrightarrow\ket{\downarrow\downarrow\uparrow}$, while processes such as $\ket{\uparrow\downarrow\uparrow}\leftrightarrow\ket{\uparrow\uparrow\uparrow}$ are forbidden. Each allowed process shifts the position of a dimer without changing the total number. At the edge of the chain, the Hamiltonian becomes $\hat{\sigma}_1^x\hat{Q}_2$ and $\hat{Q}_{N-1}\hat{\sigma}_N^x$. These terms can be understood by assuming that there is a virtual site where the atom is in the ground state (Fig.~\ref{fig:cluster}). 
\begin{figure}[htbp]
    \centering
    \includegraphics[width=1\linewidth]{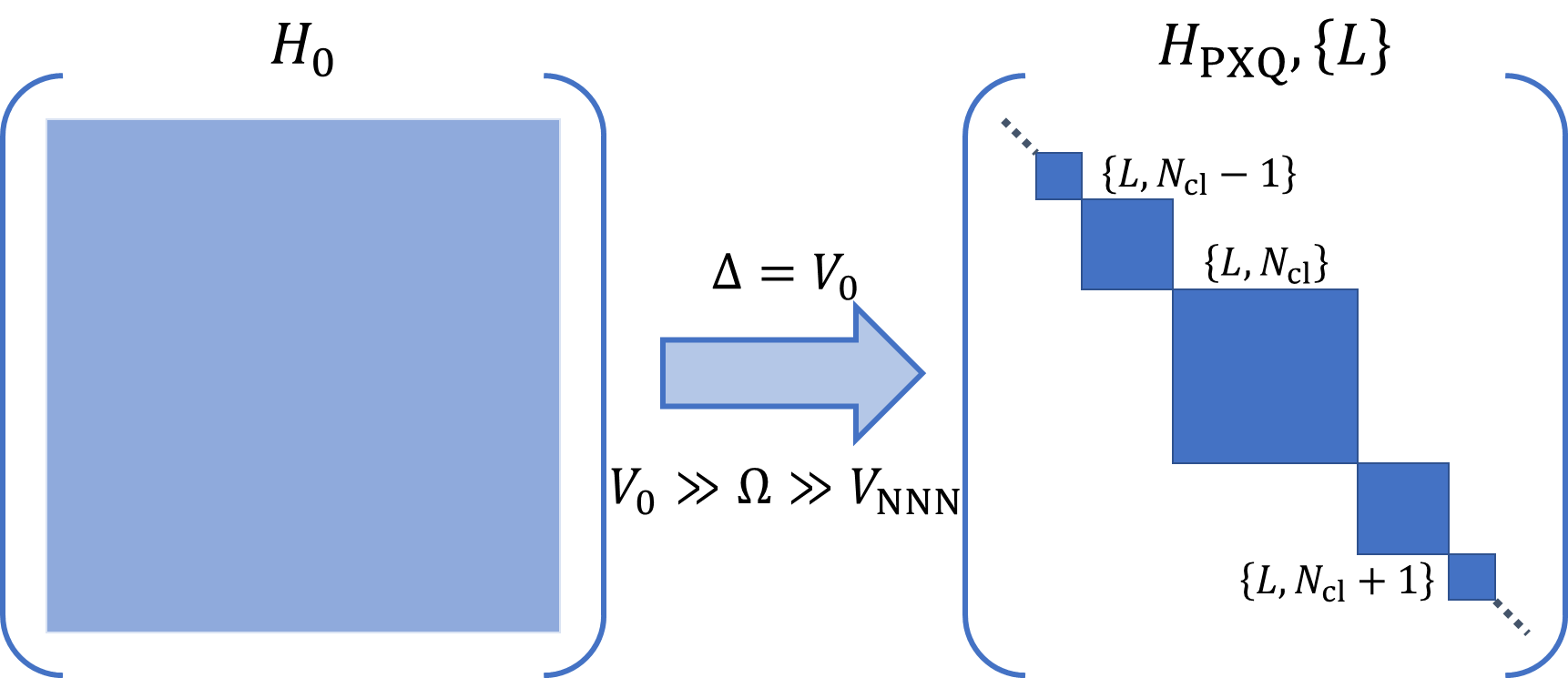}
    \caption{Schematic illustration of the block-diagonal structure of the Hamiltonian. Under the conditions $\Delta = V_0$ and $V_0 \gg \Omega \gg V_{\text{NNN}} $, the Hamiltonian $\hat{H}_0$ is effectively described by the PXQ Hamiltonian. Due to the conservation of the antiferromagnetic dimers, the PXQ Hamiltonian becomes block-diagonal.}
    \label{fig:block-diagonal}
\end{figure}

The number of antiferromagnetic dimers depends on the system size $L$ ($L>2$). Because the operator $\hat N_{\text{cl}}$ is a conserved quantity, we obtain a block decomposition of the Hilbert space according to the number of dimers~\cite{Sala2020Ergodicity},
\begin{equation*}
\mathcal{H}_{\{L\} }=\bigoplus_{N_{\text{cl}}=0}^{[(L+1)/2]} \mathcal{H}_{\{L, N_{\text{cl}}\}}
\end{equation*}
where $[\cdot]$ is the floor function, and $N_{\text{cl}}=0,1,\cdots$ is the eigenvalue of operator $\hat{N}_{\text{cl}}$. The Hamiltonian can be written in a block-diagonal form (Fig.~\ref{fig:block-diagonal}), where each block (subspace) has a fixed number of dimers. Among the subspaces, the maximal number of $\ket{\up\down}$ dimers is $[(L+1)/2]$, which gives the largest possible value of $N_{\text{cl}}$. Despite the Hilbert space becoming block-diagonal, this does not correspond to Hilbert space fragmentation, which requires that the number of subspaces scales exponentially with $L$~\cite{Sala2020Ergodicity}. Here the number of subspaces increases linearly with $L$, whereas the dimension of the overall Hilbert space grows exponentially. 

Within each block of the Hilbert space, different basis states are coupled, but this neither merges nor creates a dimer~\cite{Ostmann2019localization}. A computational basis state belongs to $\mathcal{H}_{\{L,N_{\text{cl}}\}}$ whenever the chain contains $N_{\text{cl}}$ clusters of consecutive $\ket{\uparrow}$ spins, each bounded by left- and right-edge dimers $\boxed{\downarrow\uparrow}$ and $\boxed{\uparrow\downarrow}$, as illustrated in Fig.~\ref{fig:cluster}. If one defines the spin operator by swapping operator $\hat{P}$ and $\hat{Q}$, i.e. $\tilde N_{\text{cl}}=\sum_k \hat{P}_k \hat{Q}_{k+1}$, the number of configurations is same with $\hat{N}_{\text{cl}}$. This means that the dimension of $\mathcal{H}_{\{L,N_{\text{cl}}\}}$ reduces to the number of ways to choose the $2N_{\text{cl}}$ cluster boundaries among the $L+1$ possible dimer positions, leading to the total configurations given by $N^L_{\text{cl}}=C_{L+1}^{2N_{\text{cl}}}$. By summing all the subspaces formed by different numbers of dimers, we obtain the total dimension $2^L$ of the Hilbert space.

The number conservation of the dimer indicates that the PXQ model is integrable in the thermodynamic limit. To illustrate this explicitly, we note that under the Kramers–Wannier transformation~\cite{Ostmann2019localization},
\begin{align*}
\hat{\sigma}_k^x &= \hat{\mu}_k^x \hat{\mu}_{k+1}^x,\\
\hat{\sigma}_k^y &= (-1)^{k+1}\!\left(\prod_{l=1}^{k-1}\hat{\mu}_l^z\right)\hat{\mu}_k^y\hat{\mu}_{k+1}^x,\\
\hat{\sigma}_k^z &= (-1)^{k+1}\!\left(\prod_{l=1}^{k}\hat{\mu}_l^z\right),
\end{align*}
the PXQ Hamiltonian can be mapped onto the spin-$1/2$ XX model given by,
\[
\hat{H}_{\text{PXQ}}
\;\longrightarrow\;
\hat{H}_{XX}= \frac{\Omega}{2}\sum_{k=1}^{L} 
\left(\hat{\mu}_k^x \hat{\mu}_{k+1}^x+\hat{\mu}_k^y \hat{\mu}_{k+1}^y\right),
\]
expressed in a new two-spin basis $\{\ket{0},\ket{1}\}$ via  
$\ket{\uparrow\uparrow},\ket{\downarrow\downarrow}\!\to\!\ket{0}$ and 
$\ket{\uparrow\downarrow},\ket{\downarrow\uparrow}\!\to\!\ket{1}$~\cite{Yang2019Quantum, Kim2024Realization}. The spin XX chain is one of the paradigm models in the study of low dimensional quantum magnetism~\cite{liebTwoSolubleModels1961, katsuraStatisticalMechanicsAnisotropic1962, barouchStatisticalMechanics$mathrmXY$1970}. As an integrable model, it can be solved exactly through the Jordan-Wigner transformation, which maps the ``effective spins" to free fermions.

\subsection{Coupling graph of the subspace}
\label{sec:coupling}

\begin{figure}[htbp]
    \centering
    \includegraphics[width=1.0\linewidth]{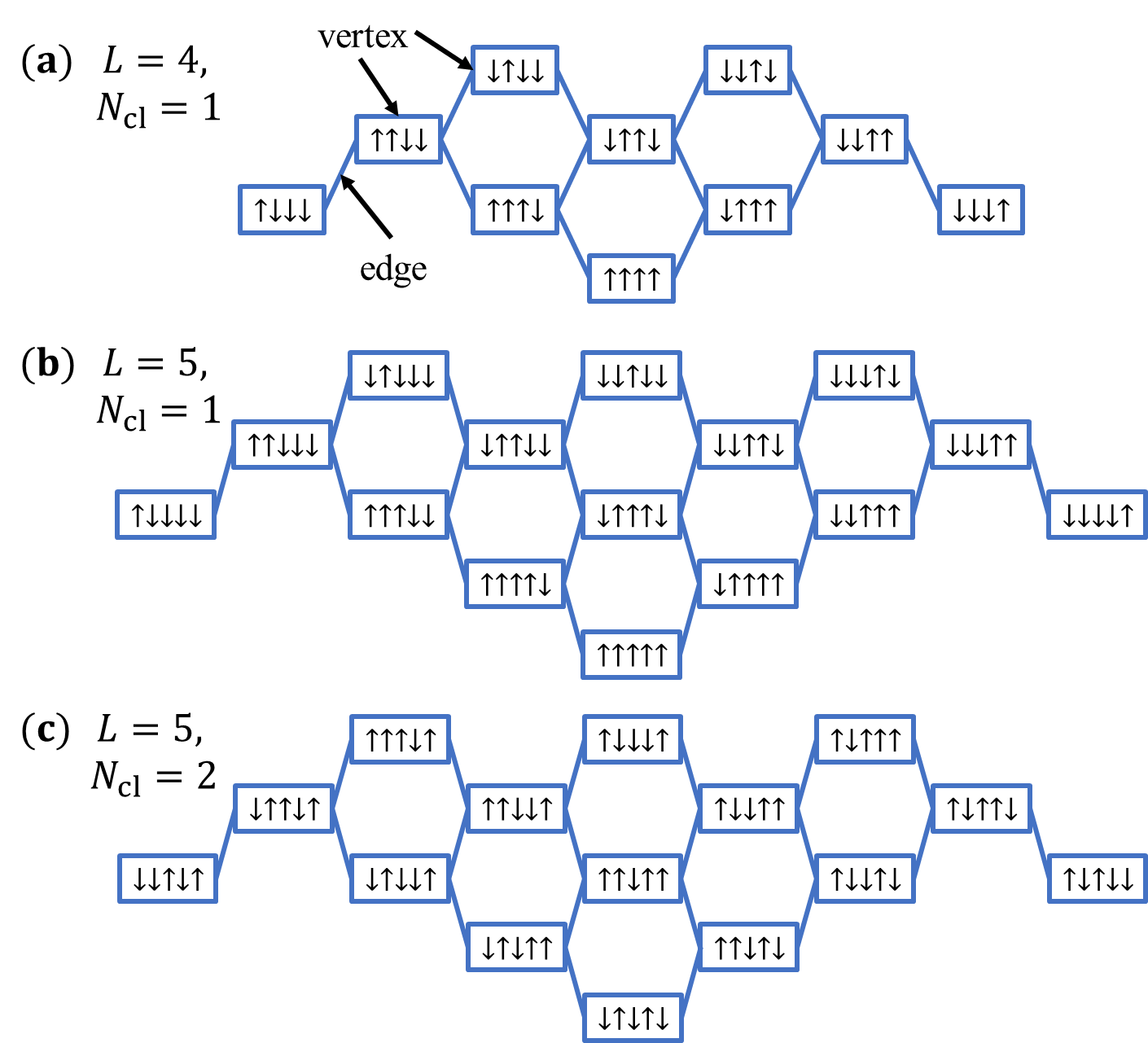}
    \caption{Coupling graphs in subspace $N_{\text{cl}} = 1$ for $L=4$(a) and $L=5$(b), and $N_{\text{cl}} = 2$ for $L=5$(c).  In the graph, each basis forms a vertex, whereas the coupling that links the neighboring basis states gives the edge. } 
    \label{fig:Ncl=1}
\end{figure}

Given a subspace, we can draw a coupling graph that shows how the dimer moves between the basis states. The sector with $N_{\text{cl}}=1$ has been widely studied~\cite{Igor2017Facilitation,Ostmann2019localization,Van2021disorder}, which admits analytical form for both the eigenvalues and eigenstates~\cite{Van2021disorder}. For this subspace, by analyzing how the three-site operators act on the left and right cluster boundaries, we obtain a coupling graph with a Pascal triangle structure~\cite{Igor2017Facilitation}. In Fig.~\ref{fig:Ncl=1}, we show the coupling graph for $L=4$ and $L=5$. The dimer shifts one site for any two neighboring basis. In each case, the graph encodes the basis states (vertices) and allowed transitions (edges). The number of vertices equals the subspace dimension $C_{L+1}^{2N_{\text{cl}}}$, whereas the number of edges corresponds to the nonzero matrix elements in the Hamiltonian of the subspace. 

For different $L$, one finds that $C_{L+1}^{2N_{\text{cl}}}=C_{L+1}^{L+1-2N_{\text{cl}}}$. When $L$ is an odd number $L=2K-1$ ($K$ is an integer),  both $2N_{\text{cl}}$ and $L+1-2N_{\text{cl}}$ are even numbers. Thus the subspaces $\mathcal{H}_{\{L,N_{\text{cl}}\}}$ and 
$\mathcal{H}_{\{L,(L+1)/2 - N_{\text{cl}}\}}$ share identical coupling structures, i.e. the subspaces with $N_{\text{cl}}$ and $K-N_{\text{cl}}$  dimers  have the same internal structure. This can also be obtained from Hamiltonian $\hat{H}_{\text{XX}}$. If we exchange the spin states $0\leftrightarrow1$ in the basis states, $\ket{\cdots 0110010\cdots}\;\longrightarrow\;\ket{\cdots 1001101\cdots}$, Hamiltonian $\hat{H}_{XX}$ is invariant.  
 In Fig.~\ref{fig:Ncl=1}(b) and (c), we illustrate this symmetry with $L=5$ and $K=2$. Both the $N_{\text{cl}}=1$ and $N_{\text{cl}}=2$ subspaces have identical coupling graphs. 

\begin{figure}[htbp]
    \centering
    \includegraphics[width=0.9\linewidth]{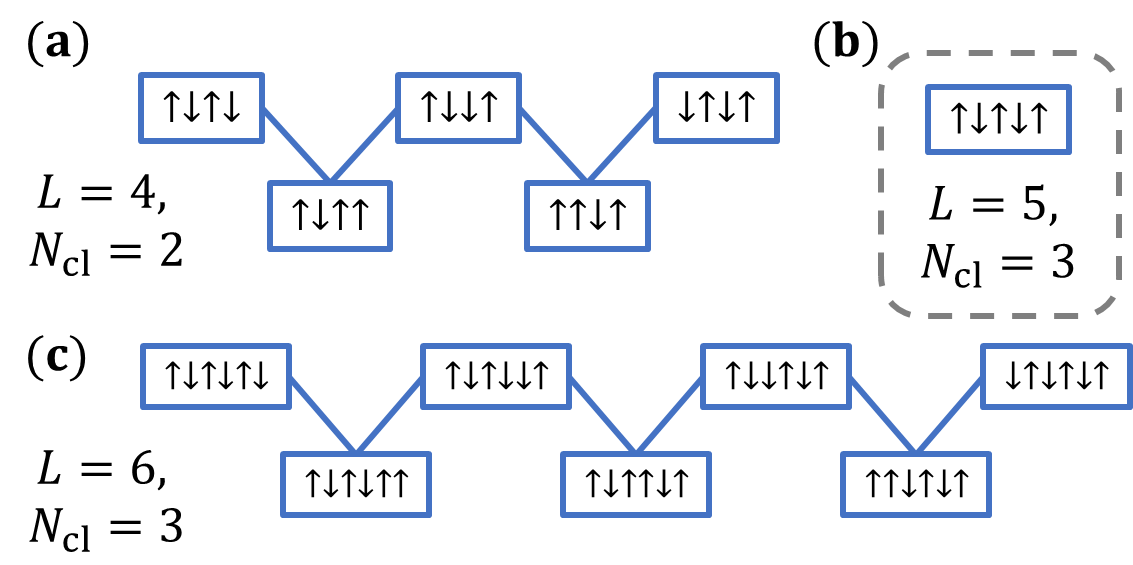}
    \caption{Coupling graph of basis states. For a given $L$, the  maximal value of $N_{\text{cl}}$ is $ = [(L+1)/2]$.  For $L=4$(a), $5$(b), and $6$(c), the corresponding ones are $2$, $3$ and $3$.}
    \label{fig:Nclmax}
\end{figure}

We now discuss the properties of the largest dimer space. When $N=2K$ is an even number, the maximal number of dimers is $N_{\text{cl}}=K$.  This subspace corresponds to the $\mathbb{Z}_2$ configuration. Only one chain end is dynamically active, leading to a linear chain of $L+1$ coupled states. We illustrate this with $L=4$ and $L=6$ in Fig.~\ref{fig:Nclmax}. The coupling graph shows that the $\mathbb{Z}_2$ pattern consists of inert domains whose interiors remain static, whereas dynamical evolution occurs only at the domain edges whenever configurations $\ket{\uparrow\uparrow\downarrow}$ or $\ket{\downarrow\downarrow\uparrow}$ appear. The resulting dynamics closely mirrors the motion of boundaries in the single-cluster case. For odd $L$, both chain ends host same spin states, giving configurations $\ket{\uparrow\downarrow\uparrow}$ and $\ket{\downarrow\uparrow\downarrow}$. Because this gives a singlet state of the PXQ Hamiltonian, the corresponding basis state is also a frozen state. We show one example with $L=5$ in Fig.~\ref{fig:Nclmax}(b). 

\section{Dimer Dynamics in the PXQ model and Rydberg atom array}
\label{Dynamics in fragmental subspaces}
The PXQ model provides an effective description of the one-dimensional Rydberg atom chain. This approximation is valid when the detuning aligns with the NN interaction $\Delta = V_{\text{0}}$ and the hierarchy of energy scales $V_{\text{0}} \gg \Omega \gg V_{\text{NNN}}$ is satisfied. In this regime, the system Hamiltonian exhibits a block-diagonal structure, where the number of dimers $N_{\text{cl}}$ is conserved. Due to the integrability, the model is characterized by nontrivial dynamical phenomena, including nonthermal behavior and the conservation of dimer density.
However, when deriving the PXQ model, the fast-rotating terms and longer-range (than the NN) interactions have been neglected. While the diagonal nature of the two-body interaction prevents the mixing of basis states, the inclusion of long-range interactions introduces state-dependent energy shifts. Importantly, the fast-rotating terms induce coupling between disparate Hilbert subspaces, breaking the conservation of dimers. Such perturbations can lead to significant deviations from the idealized dynamics when Rydberg chains are used to probe dimer properties. In this section, we investigate the roles played by these terms and identify the parameter regimes in which the underlying dynamics can be  approximately mapped to the PXQ model.

\subsection{\texorpdfstring{Single dimer subspace}{Ncl=1}}
\begin{figure}[htbp]
    \centering
    \includegraphics[width=0.94\linewidth]{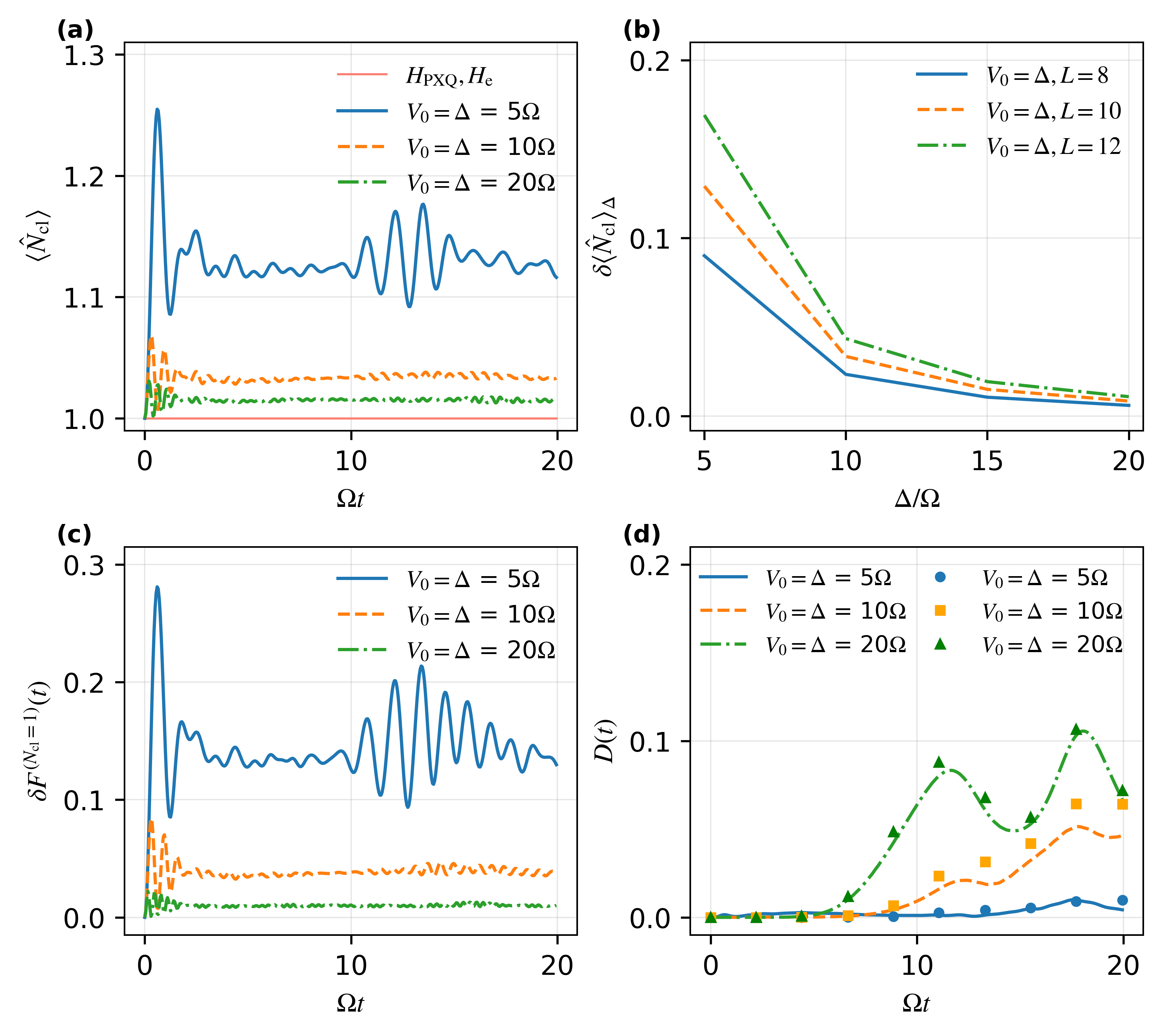}
    \caption{Dynamics in the single dimer subspace. (a) $\langle \hat{N}_{\text{cl}}\rangle$ for different $V_0$. (b) Deviations $\delta \langle\hat{N}_{\text{cl}}\rangle_0$ from the PXQ Hamiltonian. (c) Leakage of the Hilbert space from the single dimer subspace. (d) Averaged Rydberg population difference $D(t)$. Simulations with Hamiltonians $\hat{H}_0$ (lines) and $\hat{H}_e$ (labels) yield nearly identical data, indicating that the longer-range interactions cause the deviation. In all panels, we compare the dynamics {of the initial state $\ket{\psi_0}=\ket{\uparrow\downarrow\downarrow\cdots}$} driven by Hamiltonian $\hat{H}_0$ with those from $\hat{H}_{\text{PXQ}}$. In (a), (c) and (d), $L=10$. }
    \label{fig:Ncl-PXQandTrue-SingleExcited}
\end{figure}
We obtain time dependent state  $|\psi(t)\rangle = \exp(-i \hat{H}t)|\psi_0\rangle$ by solving the Sch\"odinger equation from an initial state $|\psi_0\rangle$, where the Hamiltonian $\hat{H}$ is  chosen to be $\hat{H}_0$, $\hat{H}_e$ or $\hat{H}_{\text{PXQ}}$. By comparing the dynamics of the three Hamiltonians, we identify the parameter regimes in which the dynamics of the approximate Hamiltonians $\hat{H}_e$ and $\hat{H}_{\text{PXQ}}$ capture the dynamics of the dimer. 
We consider the initial state given by the basis state $\ket{\psi_0}=\ket{\uparrow\downarrow\downarrow\cdots}$, which contains a single dimer at the left edge of the chain. 
\begin{figure}[htbp]
    \centering
    \includegraphics[width=0.94\linewidth]{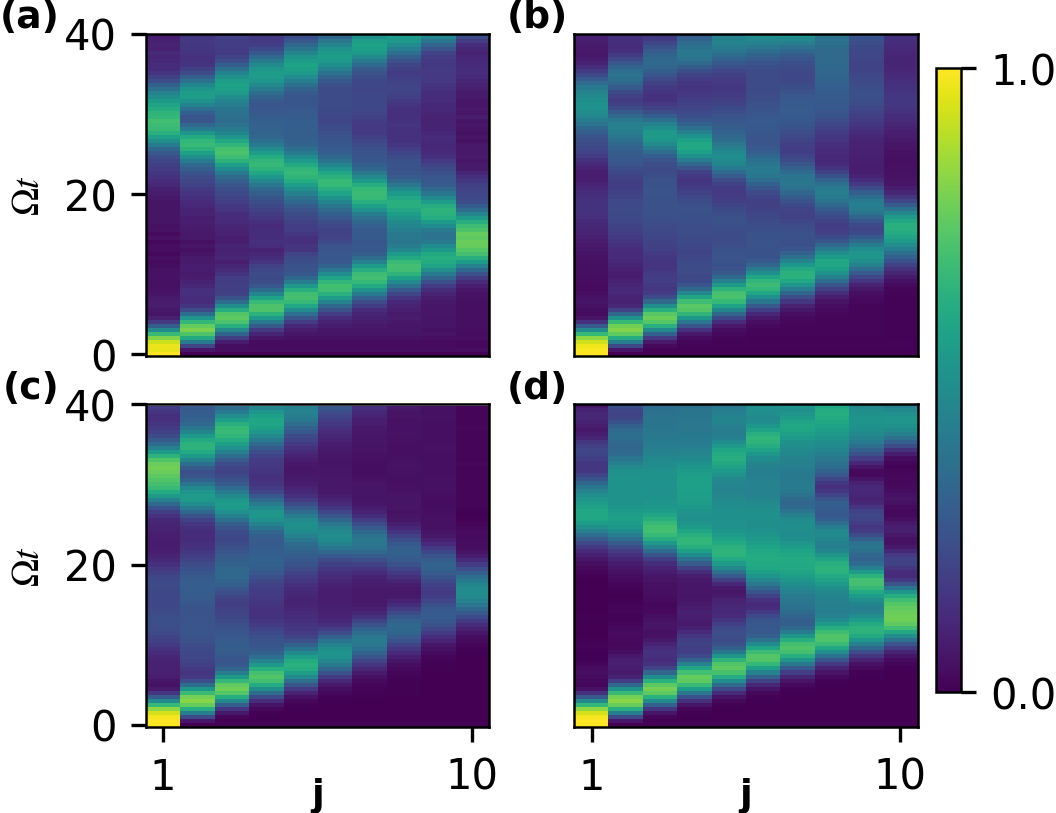}
    \caption{Dynamics of population $\langle \hat{Q}_j\rangle$ {with the initial state $\ket{\psi_0}=\ket{\uparrow\downarrow\downarrow\cdots}$}.  We show simulations for $\langle \hat{Q}_j\rangle$ for $V_0= 5\Omega$ (a), $10\Omega$ (b), and $20\Omega$ (c) with Hamiltonian $\hat{H}_0$. Data in panel (d) is obtained using $\hat{H}_{\text{PXQ}}$. In all simulations, $L =10$. }
    \label{fig:fidelity-PXQandTrue-SingleExcited}
\end{figure}

After numerically finding state $\ket{\psi(t)}$, we evaluate expectation value $\langle \hat{N}_{\text{cl}}\rangle=\bra{\psi(t)}\hat{N}_{\text{cl}}\ket{\psi(t)}$. The dynamical evolution of $\langle \hat{N}_{\text{cl}}\rangle$ for a chain with $L=10$ is shown in Fig.~\ref{fig:Ncl-PXQandTrue-SingleExcited}(a). In case of the Hamiltonians $\hat{H}_\text{PXQ}$ and $\hat{H}_\text{e}$, $\langle \hat{N}_{\text{cl}}\rangle_{\text{PXQ}}=\langle \hat{N}_{\text{cl}}\rangle_{\text{e}}=1$ because the dimer number is conserved. Here $\langle \hat{N}_{\text{cl}}\rangle_{\text{PXQ}}$ and $\langle \hat{N}_{\text{cl}}\rangle_{\text{e}}$ are the expectation values obtained with $\hat{H}_\text{PXQ}$ and $\hat{H}_\text{e}$, respectively. This is in contrast to the Hamiltonian $\hat{H}_0$, where $\langle \hat{N}_{\text{cl}}\rangle_0$ varies with time and depends on the parameters. For different $V_0$, we find that $\langle \hat{N}_{\text{cl}}\rangle_0$ fluctuates around a steady value at later times. Increasing $V_0$, the steady values gradually decrease and approach $1$. To show this trend, we evaluate the time averaged mean value $\overline{\langle \hat{N}_{\text{cl}}\rangle}_{0} =\int_0^{t_f}  \langle \hat{N}_{\text{cl}}\rangle_{0}dt$, and compare it with $\langle \hat{N}_{\text{cl}}\rangle_{\text{PXQ}}$. The difference  $\delta \langle\hat{N}_{\text{cl}}\rangle_0 = \overline{\langle \hat{N}_{\text{cl}}\rangle}_{0} -\langle \hat{N}_{\text{cl}}\rangle_{\text{PXQ}}$ is shown in Fig.~\ref{fig:Ncl-PXQandTrue-SingleExcited}(b). As $V_0$ increases, $\delta \langle\hat{N}_{\text{cl}}\rangle_0$ decreases rapidly. This tendency is largely independent of chain length $L$. This results from the fact that as $V_0$ increases, the fast-rotating terms that have been neglected in deriving the effective Hamiltonian (Appendix~\ref{app:H_e}) become less important, such that the dynamics is largely confined to the subspace in which $\hat{N}_{\text{cl}}$ is well defined. In Fig.~\ref{fig:Ncl-PXQandTrue-SingleExcited}(c), we show the leakage from the $N_{\text{cl}}=1$ subspace, $\delta F = 1 - |\bra{\psi(t)}\hat{I}_1\ket{\psi(t)}_{N_{\text{cl}=1}}|^2$, where $\hat{I}_1$ is the identity operator in the single dimer subspace. The state leakage decreases as $V_0$ increases, which is consistent with the data shown in Fig.~\ref{fig:Ncl-PXQandTrue-SingleExcited}(a)-(b).

Despite the fast-rotating terms playing marginal roles when $V_0\gg \Omega$, the long-range interactions (beyond the NN interaction) also increase. This means that the long-range interaction terms in $\hat{H}_\text{e}$ play non-negligible roles in the dynamics. In Fig.~\ref{fig:fidelity-PXQandTrue-SingleExcited}, we show the distribution of the Rydberg state $\langle \hat{Q}_j\rangle$ as a function of time, which displays the transport of dimers along the chain. When $V_0=5\Omega$, the dimer hops along the chain, and is reflected at the boundaries, as shown in Fig.~\ref{fig:fidelity-PXQandTrue-SingleExcited}(a), respectively. During propagation, the distribution gradually broadens. Increasing $V_0=10$ and $V_0=15$,  the impact of the long-range interactions is even stronger. Here the distribution $\langle \hat{Q}_j\rangle$ not only broadens, but also splits into multiple peaks (Fig.~\ref{fig:fidelity-PXQandTrue-SingleExcited}(b)-(c)). The overall profile of the dynamics is similar to that obtained from the PXQ model, as shown in Fig.~\ref{fig:fidelity-PXQandTrue-SingleExcited}(d). However, visible differences are observed when $V_0$ becomes larger.
\begin{figure}
    \centering
\includegraphics[width=0.94\linewidth]{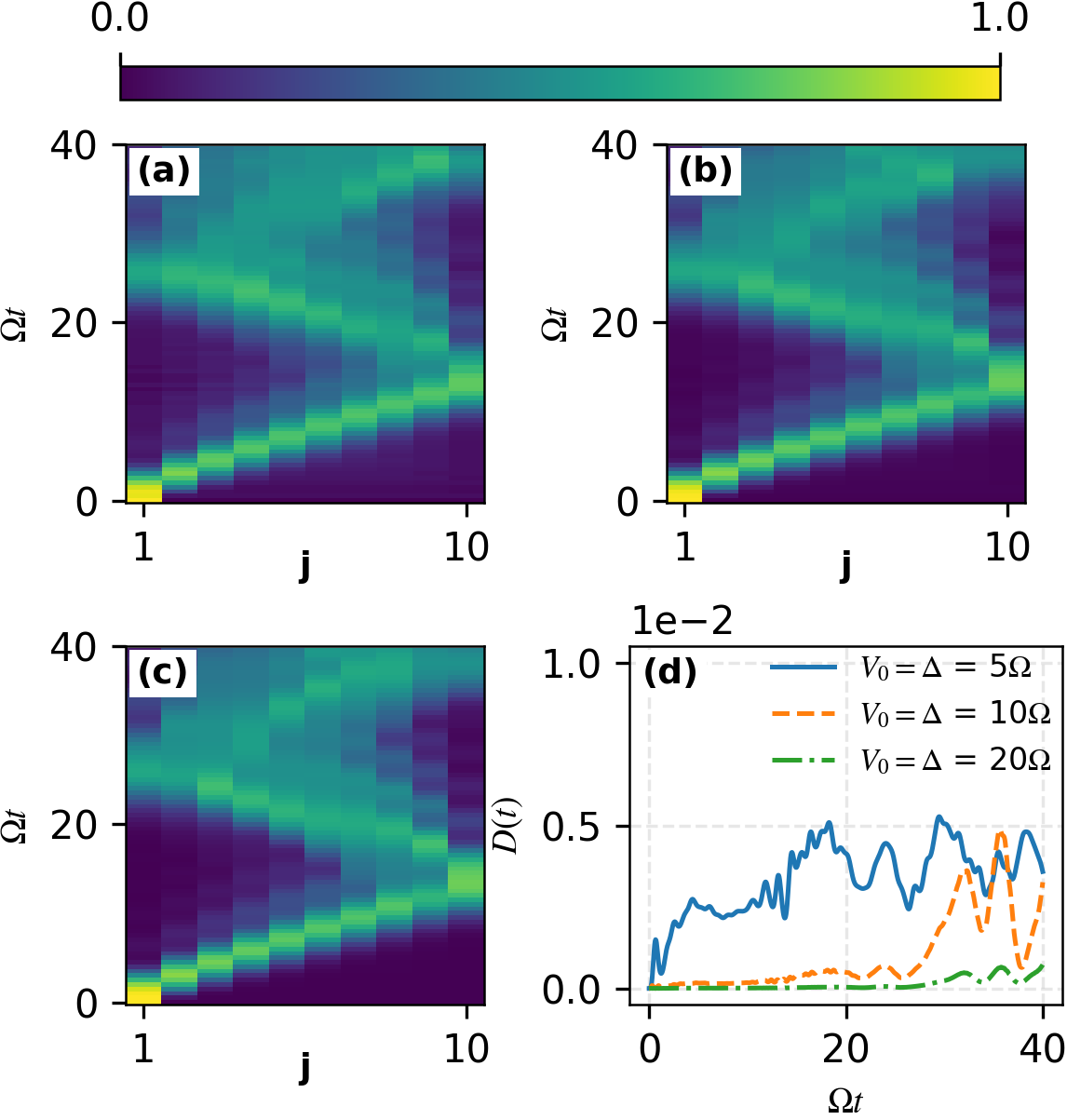}
    \caption{{Time evolution of the site-resolved excitation population for an edge single-excitation initial state at $V_0=\Delta = 5\Omega$(a), $10\Omega$(b), and $20\Omega$(c). The panel (d) shows the deviation $D(t)$ between Hamiltonian $\hat{H}_0$ with only the NN interaction and the PXQ Hamiltonian for a system of size $L = 10$.}}
    \label{fig:singletransportNN}
\end{figure}

To quantify the deviation from $\hat{H}_{\text{PXQ}}$, we calculate the overall population difference given by, 
$$D(t)=\frac{1}{L} \left( \sum_{i=1}^L |\langle \hat{Q}_i\rangle-\langle \hat{Q}_i\rangle_{\text{PXQ}}|^2 \right).$$
shown in Fig.~\ref{fig:Ncl-PXQandTrue-SingleExcited}(d). In general it becomes larger for stronger interactions. We note that the difference is primarily caused by long-range interactions. The data obtained from the Hamiltonians $\hat{H}_0$ and $\hat{H}_e$ are almost identical. To further verify this, we simulate the dynamics with $\hat{H}_0$ and keep only the NN interaction. As shown in Fig.~\ref{fig:singletransportNN}(a)-(c), the differences of the different panels from Fig.~\ref{fig:fidelity-PXQandTrue-SingleExcited}(d) are marginal, despite the inclusion of the fast-rotating terms in the simulation.  Without long-range interactions, the stronger the NN interaction, the better the agreement between the PXQ Hamiltonian and $\hat{H}_0$.  The difference in the overall population is now reduced by more than one order of magnitude, as shown in Fig.~\ref{fig:singletransportNN}(d). Such dynamics is effectively captured by the PXQ model when $V_0$ is large. It is possible to modulate profiles of the vdW interaction  by coupling multiple Rydberg states with microwave fields~\cite{marcuzziNonequilibriumUniversalityDynamics2015}. This could reduce the strength of the long-range tail, and hence minimize its impact on the dimer dynamics. 

\subsection{Maximal dimer subspace }

In this section, we study the dynamics in the subspace with the maximal number $[(L+1)/2]+1$ of dimers. When $L$ is an odd number, there is a single $\mathbb{Z}_2$  basis state $\ket{\uparrow\downarrow\uparrow\cdots\uparrow}$, which has two Rydberg atoms located at the two boundaries. As discussed in Sec.~\ref{sec:coupling}, the dynamics is frozen in this subspace. When $L=2K$ is an even number, the maximal number of dimers supported by the chain is  $N_{\text{cl}}=K $. We study the dynamics of the atom when the initial state is given by $\ket{\uparrow\downarrow\uparrow\cdots\downarrow}$. 
\begin{figure}[htbp]
    \centering
    \includegraphics[width=0.94\linewidth]{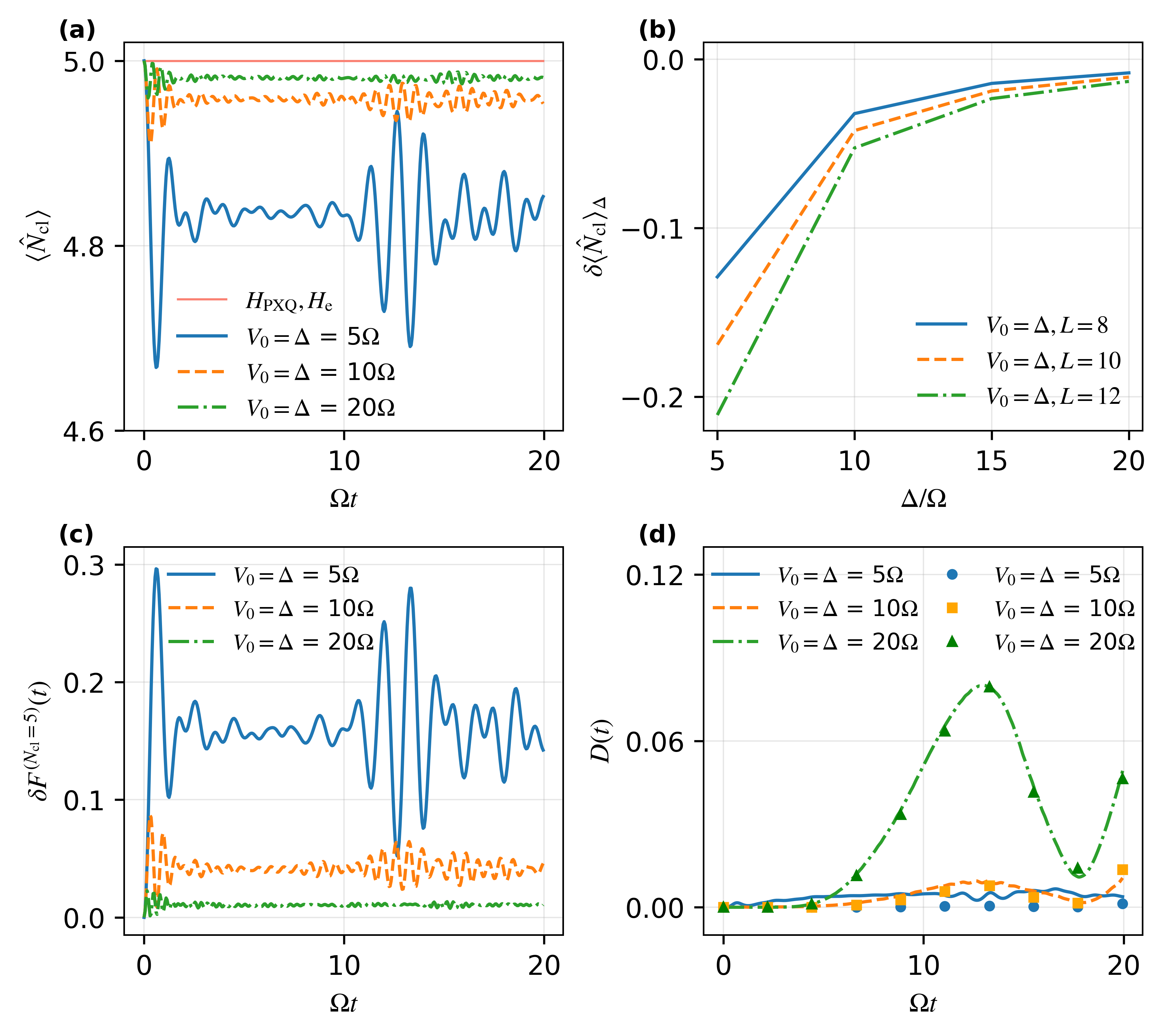}
    \caption{
    Dynamics in the maximal dimer subspace. (a) $\langle \hat{N}_{\text{cl}}\rangle$ for different $V_0$. (b) Deviation $\delta \langle\hat{N}_{\text{cl}}\rangle_0$ from the PXQ Hamiltonian. (c) Leakage of the Hilbert space from the single dimer subspace. (d) Averaged Rydberg population difference $D(t)$. Simulations with Hamiltonian $\hat{H}_0$ (lines) and $\hat{H}_e$ (labels) give nearly identical data, indicating that the longer range interactions cause the deviation. In all the panels, we compare dynamics {of the initial state $\ket{\psi_0}=\ket{\uparrow\downarrow\uparrow\cdots\downarrow}$} driven by Hamiltonian $\hat{H}_0$ with ones from $\hat{H}_{\text{PXQ}}$. In (a), (c) and (d), $L=10$. }
    \label{fig:Ncl-PXQandTrue-Z2}
\end{figure}

As shown in Fig.~\ref{fig:Ncl-PXQandTrue-Z2}(a), dynamics of $\langle N_{\text{cl}}\rangle$ is shown for different $V_0$. As $V_0$ increases, $\langle N_{\text{cl}}\rangle$ increases and approaches 5. This shows that the fast oscillating terms become less important in the dynamics. This trend exhibits good scaling with $L$. Despite the deviation increasing when $L$ is large, the trend is that it decreases with larger $V_0$, as shown in Fig.~\ref{fig:Ncl-PXQandTrue-Z2}(b). Increasing $V_0$ leads to dynamics involving basis states that are largely confined in the maximal dimer subspace (Fig.~\ref{fig:Ncl-PXQandTrue-Z2}(c)), which shows that the state leakage becomes less severe when $V_0$ increases.
 One would expect stronger effects due to the long-range interactions because the initial state contains more Rydberg atoms. In Fig.~\ref{fig:Ncl-PXQandTrue-Z2}(d), we show the difference of the Rydberg population obtained from the simulations using $\hat{H}_0$ and $\hat{H}_{\text{PXQ}}$. For $V_0=5\Omega$ and $V_0=10\Omega$, the difference is relatively small on the time scale shown in Fig.~\ref{fig:Ncl-PXQandTrue-SingleExcited}(d). It becomes apparent when $V_0=20\Omega$, but is still of the same order of magnitude when compared to the single dimer case.
\begin{figure}[htbp]
    \centering
    \includegraphics[width=0.94\linewidth]{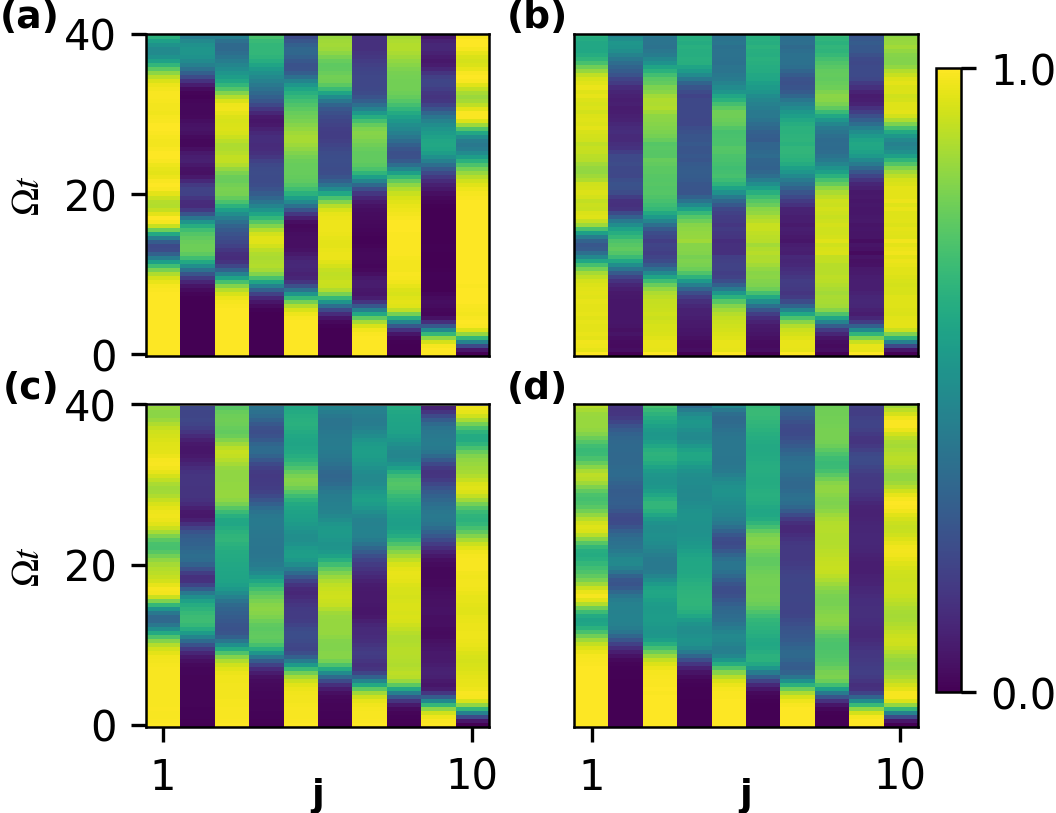}
    \caption{Dynamics of population $\langle \hat{Q}_j\rangle$ {with the initial state $\ket{\psi_0}=\ket{\uparrow\downarrow\uparrow\cdots\downarrow}$}. Data in panel (a) is obtained using $\hat{H}_{\text{PXQ}}$. We show simulations with $V_0= 5\Omega$ (b), $10\Omega$ (c), and $20\Omega$ (d) with Hamiltonian $\hat{H}_0$.  In all the simulations $L =10$.}
    \label{fig:fidelity-PXQandTrue-Z2}
\end{figure}

When considering the transport in the PXP Hamiltonian, an excitation created at the right edge propagates along the chain with a constant velocity, and is reflected when it reaches the opposite boundary,  as shown in Fig.~\ref{fig:fidelity-PXQandTrue-Z2}(a). Here dynamics is accurately described by free fermions~\cite{liebTwoSolubleModels1961,katsuraStatisticalMechanicsAnisotropic1962,barouchStatisticalMechanics$mathrmXY$1970}. When compared to $\hat{H}_0$, this picture largely holds if $V_0$ is not very strong, (Fig.~\ref{fig:fidelity-PXQandTrue-Z2}(b)).  
Further increasing the NN interaction, and hence the long-range interactions, the difference between the dynamics becomes apparent, as shown in fig.~\ref{fig:fidelity-PXQandTrue-Z2}(c)-(d). Due to the long-range interactions, it is not possible to map $\hat{H}_{\text{e}}$ to the XX Hamiltonian. In other words, one can not find a simple dispersion relation cannot be found to describe the excitation propagation. Dynamically long-range interactions introduce more frequency components, which destructively interfere with the components due to the NN interaction. Despite the visibility of $\langle \hat{Q}\rangle$ decreasing with increasing $V_0$, we emphasize that the number of dimers is conserved in the PXQ model and in $\hat{H}_0$ ($\hat{H}_e$) in the strong interaction regime.

\section{Conclusion}
\label{sec:conclusion}

Starting from a one-dimensional Rydberg atom chain, we have derived the effective model under strong interaction and single-photon resonance condition. The effective model  conserves the number of antiferromagnetic dimers $N_{\text{cl}}$. By neglecting the long-range interactions, the effective model reduces to the PXQ Hamiltonian. We have analyzed the subspace structure of the PXQ model, showing that these subspaces can be systematically classified into distinct groups. We have shown that the PXQ model captures the dynamics of the Rydberg atom chain, where the conservation of the dimer is verified through numerical simulations. Long-range interactions and fast-rotating terms cause deviations from the PXQ model. These deviations can be controlled by tuning the parameters of Rydberg atoms. Future directions include extending this study to higher-dimensional lattices, investigating its implications for quantum information applications, exploring how the symmetry can be violated by the interactions and laser coupling, and exploring dissipative effects (i.e., finite Rydberg lifetimes) when quantum simulating dimer dynamics with Rydberg atom arrays. 

\textit{Acknowledgments}--W. L. thanks Juan Garrahan and Tianyi Yan for the useful discussions. F. K. and L. L. acknowledge support the National Key Research and Development Program of China (Grant No. 2021YFA1402003), the National Science and Technology Major Project of the Ministry of Science and Technology of China (Grant No. 2023ZD0300901), the National Natural Science Foundation of China (Grant Nos. 12374329, U21A6006 and 12404580), the Science and Technology Commission of Shanghai Municipality (Grant No. 25LZ2601002). WL. acknowledges support from the EPSRC through Grant No. EP/W015641/1 and No. EP/W524402/1. The data that support the findings of this study are openly available~\cite{kuang2026Dynamics}.

\appendix

\section{Derivation of the Effective Hamiltonian }
\label{app:H_e}
The original Hamiltonian is
$$\hat{H}_0 =-\Delta \sum_{i=1}^L \hat{Q}_i +\frac{\Omega}{2} \sum_{i=1}^L \hat{\sigma}^x_i+  \sum_{i=1}^L \sum_{j=i+1}^L V_{ij}\hat{Q}_i \hat{Q}_{j},$$
where $V_{ij}$ denotes the long-distance interactions. 

For $V_0=\Delta\gg \Omega$ and using the unitary operator 
$$  \hat{U} = \exp [it\left({\sum_{j=1}^{L-1}V_0\hat{Q}_j \hat{Q}_{j+1}} -V_0 \sum_{i=1}^L \hat{Q}_i \right)  ],$$
we obtain the Hamiltonian in the interaction picture,
\begin{align*}
    \hat{H}' &= \hat{U} \cdot \left( \frac{\Omega}{2} \sum_{i=1}^L \hat{\sigma}^x_i +\sum_{i=1} \sum_{j=i+2} V_{ij}\hat{Q}_i \hat{Q}_{j} \right)\cdot \hat{U}^\dagger\\
     &=\frac{\Omega }{2} \left\{\sum_{i=1}^L \hat{\sigma}^+_i \left[ (\hat{Q}_{i-1} \hat{P}_{i+1} +\hat{P}_{i-1}\hat{Q}_{i+1}) \right. \right.\\
    &\left.\left.+e^{-iV_0 t}\hat{P}_{i-1}\hat{P}_{i+1} + e^{iV_{0}t} \hat{Q}_{i-1}\hat{Q}_{i+1} \right]+ h.c. \right\} \\
     &+\sum_{i=1} \sum_{j=i+2} V_{ij}\hat{Q}_i \hat{Q}_{j}.
\end{align*}

As $V_0=\Delta\gg \Omega$, the time-dependent terms become highly oscillatory and can be neglected.  This leads to the effective Hamiltonian,
\begin{align*}
     \hat{H}_{\text{e}}&\approx \hat{H}_{PXQ}+\sum_{i=1} \sum_{j=i+2} V_{ij}\hat{Q}_i \hat{Q}_{j}.
\end{align*}
When the long-range interaction is not important, one can neglect this term and obtain the PXQ Hamiltonian $\hat{H}_{\text{PXQ}}$.

\section{Kramers-Wannier transformation}

First, we require a transformation that satisfies the following mapping of configurations $\ket{\up\up}, \ket{\down \down} \to \ket{0}$, $\ket{\up\down}, \ket{\down\up} \to \ket{1}$, which implies that adjacent spin pairs are projected to state $\ket{0}$ if they are identical, and to state $\ket{1}$ if they are distinct. This is similar to the XOR-calculator. 
We define that $\hat{\mu}^{x,y,z}$ are Pauli matrices in the basis $\{\ket{0},\ket{1}\}$ and $\hat{\sigma}^{x,y,z}$ are Pauli matrices in the basis $\{\ket{\down},\ket{\up}\}$. 
Therefore, $\hat{Q}^\mu_j=\hat{Q}^\sigma_{j-1} \hat{P}^\sigma_j +\hat{P}^\sigma_{j-1} \hat{Q}^\sigma_j$, where $\hat{P}$ and $\hat{Q}$ denote the population number {operators} of the ground state and excited state, respectively, the subscript $\mu$ or $\sigma$ designates distinct basis sets, and the superscript $j$ indicates the corresponding lattice site position (for example, $\hat{Q}^\mu_j$ means the population {number operator} of $\ket{1}$ in $j$-th site.).
And we have $\hat{Q}^{\mu(\sigma)} =(\mathbb{I}+\mu^z(\sigma^z))/2, \hat{P}^{\mu(\sigma)} =(\mathbb{I}-\mu^z(\sigma^z))/2$.

Using the above relation, we obtain,
$$\hat{\mu}_j^z = -\hat{\sigma}_j^z \hat{\sigma}_{j-1}^z.$$
Critically, the boundary conditions require meticulous analysis, necessitating a return to the PXQ model. The PXQ model with an open boundary condition has the 2-site operators $\hat{\sigma}^x_1 \hat{Q}_2$ and $\hat{Q}_{L-1} \hat{\sigma}^x_L $, which implies virtual fixed ground state atoms in the $0$-th site and $(L+1)$-th site. If we consider the two virtual atoms in the system, the transformation is exactly trying to map a $(L+2)$-length system to a $(L+1)$-length system. 
With such assumptions we could have $\hat{\mu}_1^z = -\hat{\sigma}_1^z \hat{\sigma}_{0}^z=\hat{\sigma}_{1}^z$.
If we compute the cumulative product from 1 to k, we obtain the inverse relation,
$$ \hat{\sigma}_k^z = (-1)^{k+1} \prod_{l=1}^k \hat{\mu}_l^z.$$

The PXQ model has the couplings $ \ket{\up\up\down} \leftrightarrow \ket{\up\down\down} $ and $\ket{\down\down\up} \leftrightarrow \ket{\down\up\up}$. These transitions can be described by the "collective" spin operator $\hat{\sigma}_k^x$. In the basis $\{\ket{0},\ket{1}\}$ these couplings can be expressed by $\ket{01}\leftrightarrow\ket{10}$.  This leads to the following relation, 
$$\hat{\sigma}_k^x = \hat{\mu}_k^x \hat{\mu}_{k+1}^x.$$ 
The right-hand side also describes transitions $|00\rangle\leftrightarrow|11\rangle$, which are not present in the PXQ Hamiltonian. This type of transition is canceled by $\hat{\mu}_k^y \hat{\mu}_{k+1}^y.$ Once we have the transformation for $\hat{\sigma}_k^x$ and $\hat{\sigma}_k^z$, utilizing the commutation relations of the Pauli matrices, we obtain the transformation,
$$ \hat{\sigma}_k^y = (-1)^{k+1} \prod_{l=1}^{k-1} \hat{\mu}_l^z \hat{\mu}_k^y \hat{\mu}_{k+1}^x.$$

Then, we calculate the inverse transformation of $\hat{\mu}_k^x$. We compute the cumulative product of $\hat{\sigma}_k^x$ from $1$ to $(j-1)$ and find,
$$\hat{\mu}_j^x = \hat{\mu}_1^x \prod_{l=1}^{j-1} \hat{\sigma}_l^x.$$
Through the commutation relations of the Pauli matrices, we obtain 
$$\hat{\mu}_j^y = -\hat{\mu}_1^x \prod_{l=1}^{j-2} \hat{\sigma}_l^x \hat{\sigma}_{j-1}^y \hat{\sigma}_{j}^z.$$ With that, the inverse Kramers-Wannier transformation is obtained,
\begin{align*}
    \hat{\mu}_j^x &= \hat{\mu}_1^x \prod_{l=1}^{j-1} \hat{\sigma}_l^x, \\
    \hat{\mu}_j^y &= -\hat{\mu}_1^x \prod_{l=1}^{j-2} \hat{\sigma}_l^x \hat{\sigma}_{j-1}^y \hat{\sigma}_{j}^z, \\
    \hat{\mu}_j^z &= -\hat{\sigma}_j^z \hat{\sigma}_{j-1}^z, \quad \hat{\mu}_1^z = \hat{\sigma}_1^z.
\end{align*}

\bibliography{ref}
\end{document}